\documentclass{aastex}
\usepackage{emulateapj5,psfig}
\usepackage{graphicx}
\usepackage{timesfonts}
\makeatletter
{\end{center}\end{minipage}\smallskip}

\newenvironment{inlinefigure}{%
\def\@captype{figure}%
\noindent\begin{minipage}{0.999\linewidth}\begin{center}}
{\end{center}\end{minipage}\smallskip}
\makeatother
\def\ltsima{$\; \buildrel < \over \sim \;$}
\def\lsim{\lower.5ex\hbox{\ltsima}}
\def\loe{\lower.5ex\hbox{\ltsima}}
\def\gtsima{$\; \buildrel > \over \sim \;$}
\def\gsim{\lower.5ex\hbox{\gtsima}}
\def\goe{\lower.5ex\hbox{\gtsima}}

\newcommand{\be} {\begin{equation}}

\newcommand{\ee} {\end{equation}}

\def\uu {4U\,0142$+$614}
\def\car {1E\,1048.1$-$5937}

\def\src {1RXS J170849$-$400910}
\def\ee {1E\,2259$+$586}
\newcommand{\AXAF}{{\em Chandra}}
\newcommand{\R}{{\em ROSAT}}
\newcommand{\A}{{\em ASCA}}

\newcommand{\BSAX}{{\em Beppo}SAX} 
\newcommand{\RXTE}{{\em R}XTE}
\newcommand{\bc}{\begin{center}}
\newcommand{\ec}{\end{center}}
\def\ltsima{$\; \buildrel < \over \sim \;$}
\def\lsim{\lower.5ex\hbox{\ltsima}}
\def\loe{\lower.5ex\hbox{\ltsima}}
\def\gtsima{$\; \buildrel > \over \sim \;$}
\def\gsim{\lower.5ex\hbox{\gtsima}}
\def\goe{\lower.5ex\hbox{\gtsima}}
\newcommand {\rc}{\rm}

\lefthead{ISRAEL ET AL.}
\righthead{THE IR COUNTERPART TO THE AXP \src}
\submitted{Received: 19 November 2002;~~ Accepted 11 April 2003}

\begin{document}

\title{THE IR COUNTERPART TO THE ANOMALOUS X--RAY PULSAR
\src\altaffilmark{1}}

\authoremail{gianluca@mporzio.astro.it}

\author{G.L. Israel\altaffilmark{2,3}, S. Covino\altaffilmark{4},
R. Perna\altaffilmark{5}, R. Mignani\altaffilmark{6},
L. Stella\altaffilmark{2,3}, S. Campana\altaffilmark{4,3},
G. Marconi\altaffilmark{7}, G. Bono\altaffilmark{2},
S. Mereghetti\altaffilmark{8}, C. Motch\altaffilmark{9}
I. Negueruela\altaffilmark{10}, T. Oosterbroek\altaffilmark{11}, and
L. Angelini\altaffilmark{12}}

\footnote{The results reported in this Letter are partially based on 
observations carried out at ESO, La Silla, Chile (63.H--0294, 66.D-0440, 
67.D-0116 and 68.D-0350), and the Canada--France--Hawaii 
Telescope operated by the National Research Council of Canada, the
Centre National de la Recherche Scientifique de France and the
University of Hawaii (02bf21).}

\affil{2. INAF -- Osservatorio Astronomico di Roma, V. Frascati 33, 
       I--00040 Monteporzio Catone (Roma), 
       Italy; gianluca@mporzio.astro.it}

\affil{3. Affiliated to the International Center for Relativistic 
Astrophysics}

\affil{4. INAF -- Osservatorio Astronomico di Brera, Via Bianchi 
46, I--23807 Merate (Lc), Italy}

\affil{5. Harvard-Smithsonian Center for Astrophysics, 
60 Garden Street, Cambridge, MA 02138, USA}

\affil{6. European Southern Observatory, Karl--Schwarzschildstr. 2, 
D--85748 Garching, Germany}

\affil{7. European Southern Observatory, Casilla 19001, Santiago, 
Chile}

\affil{8. CNR--IASF, Istituto di Astrofisica Spaziale 
e Fisica Cosmica, Sezione di Milano ''G.Occhialini'', 
Via Bassini 15, I--20133 Milano, Italy}

\affil{9. Observatoire de Strasbourg, 11, rue de l'Universite, 
67000 Strasbourg, France}

\affil{10. Dpto. de F\'{\i}sica, Ingenier\'{\i}a de Sistemas y 
Teor\'{\i}a de la Se\~{n}ales, Universidad de Alicante, Apdo. de Correos 
99, E03080, Alicante, Spain}

\affil{11. Space Science Department of ESA, 
ESTEC, P.O. Box 299, 2200 AG Noordwijk, The Netherlands}

\affil{12. Laboratory of High Energy Astrophysics, Code 660.2,
NASA/Goddard, Space Flight Center, MD 20771, USA}

\begin{abstract}
We report the discovery of the likely IR counterpart to the Anomalous
X--ray Pulsar (AXP) \src, based on the combination of \AXAF\ HRC--I
X--ray position, and deep optical/IR observations carried out from ESO
telescopes and the Canada France Hawaii Telescope (CFHT) during
1999--2002.  Within the narrow uncertainty region we found two
relatively faint ($K'$=20.0 and $K'$=17.53) IR objects.  Based on
their color and position in the $J$-$K'$ versus $J$-$H$ diagram only
the brighter object is consistent with the known IR properties of the
counterparts to other AXPs. No variability was detected for this
source, similarly to what is observed in the case of \uu. {\rc Like in
other AXPs, we found that the IR flux of \src\ is higher then expected
for a simple blackbody component extrapolated from the X--ray
data}. If confirmed, this object would be the fourth IR counterpart to
a source of the AXP class, and would make the IR excess a likely new
characteristic of AXPs.

\end{abstract}


\section{INTRODUCTION} 
{\rc AXPs are thought to be solitary magnetic rotating neutron stars
(NSs) either with a standard magnetised field or larger than 10$^{14}$
Gauss (magnetars), although the binary system scenario with a very low
mass companion is not completely ruled out by current observational
data. (for a review see Israel et al. 2002a; Mereghetti et al. 2002
and references therein)}.  Different production mechanisms for the
observed X--ray emission have been proposed, involving either
accretion or dissipation of magnetic energy.  {\rc The magnetar model,
originally proposed by Duncan \& Thompson (1992) to explain Soft
Gamma--ray Repeaters (SGRs), appears to be successful at interpreting
most of the properties of AXPs.}  In fact, AXPs have been linked to
SGRs (thought to be magnetars, neutron stars powered by their strong
magnetic fields) because of similar timing properties, namely periods
and positive period derivatives (Duncan \& Thompson 1992; Thompson \&
Duncan 1993; 1996).  What differentiates the two classes of objects is
unclear. The similarity in the spin parameters would not be sufficient
by itself to differentiate AXPs and SGRs from other groups of pulsars
with very different emission properties. Conversely, a very high
magnetic field strength (if at all) cannot be the sole factor
governing whether or not a neutron star is a magnetar, a radio pulsar
or in a binary system (Camilo et al. 2000).

The possible connection of AXPs with SGRs has gained more credibility
with the recent detection of SGR--like X--ray bursts from two AXPs
(Kaspi \& Gavriil 2002, Gavriil et al. 2002) which also showed IR
variability. 
For \ee, IR variability of the likely counterpart has been
detected few days after a strong bursting activity seen with the
\RXTE\ (Kaspi et al. 2002, Israel et al. 2003). Similarly, the
variability of the IR counterpart to \car\ is thought to be related to
X--ray variability (Israel et al. 2002b), though no simultaneous
X--ray/IR observations were available.  The magnetar scenario does not
make any prediction (yet) on the optical/IR emission seen in AXPs. If
AXPs are powered by accretion via a disc, in analogy with the X--ray
bursts observed in Low Mass X--ray Binaries, it is expected that the
effects of the X--ray bursts activity in AXPs (and SGRs) propagate
toward longer wavelengths e.g., via reprocessing in the disc (see
Middleditch \& Nelson 1976; Lawrence et al. 1983). Evidence
for flattening {(or excess)} of the flux distribution in the IR band
{\rc (with respect to a simple}
\begin{figure*}
\psfig{figure=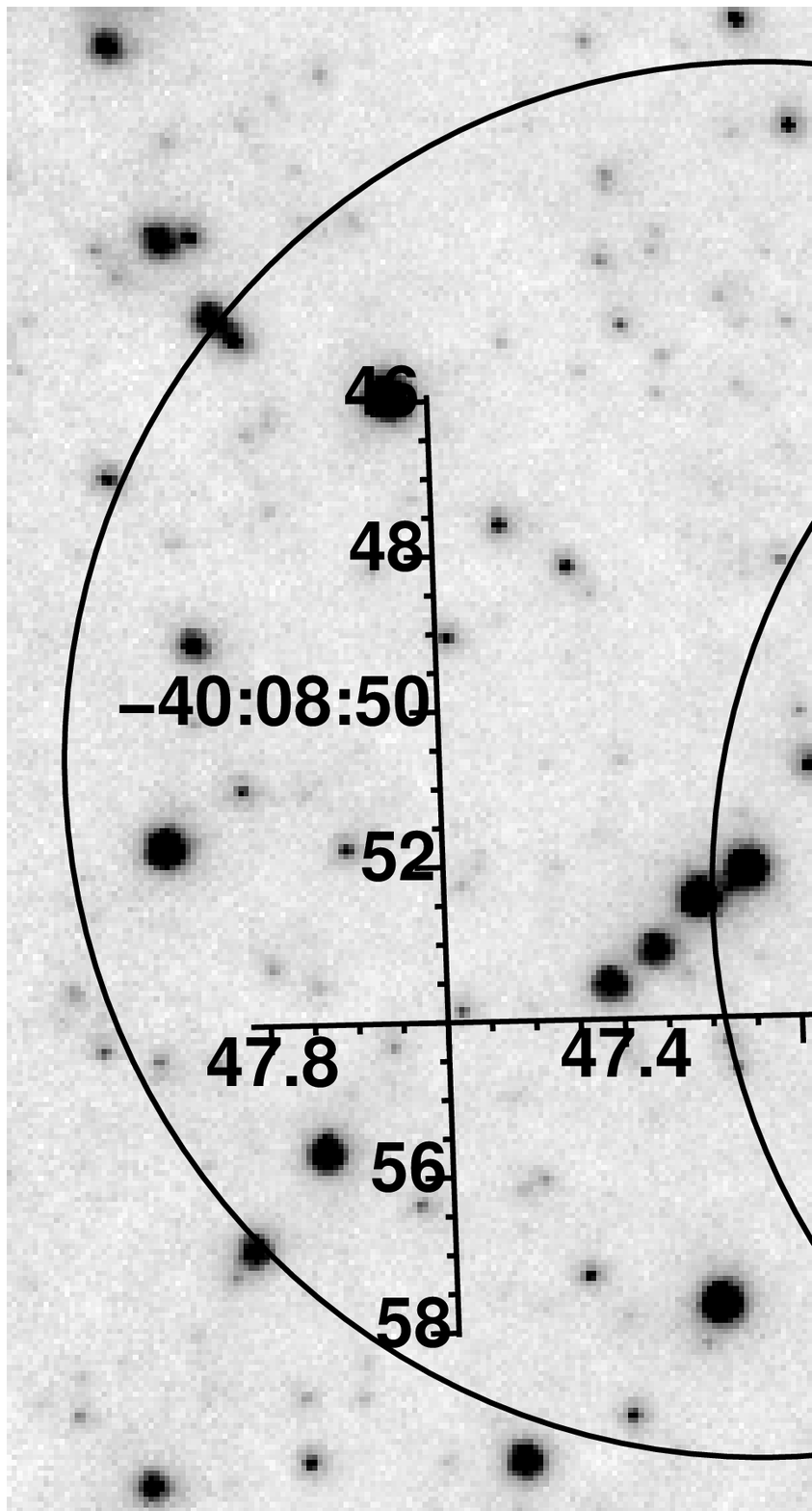,height=4.6cm}
\psfig{figure=RXSzKpne.ps,height=4.6cm} 
\caption{ Near--IR $K'$ band image of the
region including the position of \src, taken from the 3.6-m CFHT. The
\R\ and \AXAF\ uncertainty circles (9\arcsec\ and 0.8\arcsec,
respectively) are also shown.  The inset shows the close--up of the
\AXAF\ position with the two counterpart candidates marked (letter A and
B).  Coordinates are RA (h\,m\,s) and Dec. ($^o$ \arcmin\ \arcsec;
equinox J2000).}
\end{figure*}
{ blackbody component forced to be
consistent both with the X--ray and optical data)} has been reported
in three AXPs, namely \ee, \car\ and \uu\ (Hulleman et al. 2001; Wang
\& Chakrabarty 2002; Israel et al. 2003). The spectral flattening and
variability of the IR emission of AXPs are a potentially important
diagnostics for the study of these enigmatic objects and their
possible connections with other classes of pulsars.

\src\ was discovered by the \R\ mission during the All Sky Survey
program (Voges et al. 1996), but only a few years later $\sim$11\,s
pulsations were discovered by \A\ (Sugizaki et al. 1997). Both the
period derivative (\.P$\sim$2$\times$10$^{-11}$ s\,s$^{-1}$) and the
optical upper limits inferred for its counterpart were found to be
consistent with those of AXPs (Israel et al. 1999). The source is
radio quiet (no pulsations detected in the radio band; Israel et
al. 2002a) and a stable rotator comparable to radio pulsars (Gavriil
\& Kaspi 2002).  It is also the only AXP for which X--ray spectral
variations as a function of the pulse phase have been detected
associated to variations of the pulse profile (similar to those
observed in ``standard'' accretion--powered X--ray binary systems;
Israel et al. 2003).  Finally, \src\ represents the only AXP for which
a spectral signature, thought to be due to resonant cyclotron
absorption, has been detected confirming the presence of a relatively
high magnetic field (Rea et al. 2003).

In this paper, we present the results obtained from \AXAF, ESO and
CFHT observations of \src\ in the X--ray, optical and IR bands.  We
identified the likely IR counterpart to this AXP based on positional
coincidence and unusual IR colors. The overall (from X--ray to IR)
energy distribution is also presented and discussed in the light of
the {\rc IR excess detected in \src\ and other AXPs.}

\section{\AXAF\ OBSERVATIONS}

The field of \src\ was observed during \AXAF\ Cycle 2 with the High
Resolution Imager (HRC--I; Zombeck et al. 1995) on 2001 September 23
for an effective exposure time of 9870\,s. Data were reduced with CIAO
version 2.2 and analysed with standard software packages for X--ray
data (CIAO, XIMAGE, XRONOS, etc.). The observation was carried out
with a nominal aspect solution and the latest calibration files were
used. Only one source was detected in the HRC--I (see Israel et
al. 2002b for details on the detection algorithms). The source has the
following coordinates: R.A. = 17$^{\rm h}$08$^{\rm m}$46\fs87, DEC. =
--40\arcdeg08\arcmin52\farcs44 (equinox J2000), with an uncertainty
circle radius of 0\farcs7 (90\% confidence level; consistent with \R\
HRI positions in Israel et al. 1999). Photon arrival times were
extracted from a circular region with a radius of 1\farcs5, including
more than 90\% of the source photons, and corrected to the barycenter
of the solar system. A coherent pulsation at a period of
11.0011$\pm$0.0005\,s (90\% confidence level) was detected confirming
that the source is indeed \src.

We also carried out a spatial profile analysis for the (background
subtracted) X--ray emission from a region of 7\arcsec\ radius centered
on the above reported source position. The spatial profile was found
to be in good agreement with the expected \AXAF\ Point Spread Function
(PSF) for an on--axis source (see Israel et al. 2002b for details).
Unfortunately, the high X--ray background level detected in the \src\
field prevented a more sensitive search for diffuse emission at larger
radii.

\section{OPTICAL/IR OBSERVATIONS} 

Most of the optical/IR data were obtained from the 3.6-m ESO telescope
(ESO360; La Silla, Chile) equipped with {\tt ESO Faint Object
Spectrograph and Camera} (EFOSC2; a CCD with 2048$\times$2048 pixels;
0\farcs157 pixel scale and 5.2\arcmin$\times$5.2\arcmin\ field of
view, FOV) in the optical band, and with the 3.5-m New Technology
Telescope (NTT; La Silla, Chile)
\begin{inlinefigure}
\psfig{figure=rxs_cce.ps,height=7.9cm} 
\caption{Color--color diagram obtained for
the objects detected within a radius of 30\arcsec\ around
the \AXAF\ position of \src. The proposed counterpart is marked (A). 
Star sequences are shown for comparison, for different values of 
$A_{\rm V}$: 0, 2, 4, 6, and 11.2 (thin solid lines; the latter value 
obtained for the total Galactic absorption), and $A_{\rm V}$=7.8 
(thick solid line; assuming $N_{\rm H}$=1.4$\times$10$^{22}$ cm$^{-2}$ 
as inferred from X--ray observations).  The rectangle which includes the 
position of object A represents the most probable region of the diagram 
where the IR counterpart to \src\ would lie, based on the known IR 
counterparts to AXPs.}
\end{inlinefigure}

\noindent 
equipped with the {\tt Son OF Isaac} (SOFI; Hawaii HgCdTe
1024$\times$1024 array; 0\farcs292 pixel scale and
4.9\arcmin$\times$4.9\arcmin\ FOV) in the near--IR band. Additional IR
data were obtained at the 3.6-m CFHT (Mauna Kea, Hawaii) equipped with
the {\tt Adaptive Optics Bonnette} (AOB; Hawaii HgCdTe
1024$\times$1024 array; 0\farcs035 pixel scale and
36\arcsec$\times$36\arcsec\ FOV).

Optical Johnson $R$ filter deep images were obtained with {\tt EFOSC2}
on 1999 September 15--16 with effective exposure times of 1500s
(seeing of 0\farcs8). Standard reduction packages were used in the
analysis of the optical data ({\tt DAOPHOT\,II}; Stetson 1987) in
order to obtain the photometry of each stellar object in the images. A
limiting magnitude (3$\sigma$ confidence level) of $R$$\sim$26.5 was
reached.

Images in the $J$ and $H$ bands were initially acquired with {\tt
SOFI} on 2001 May 26 with 1920\,s ($J$) and 2400\,s ($H$) of effective
exposure time (seeing of 0\farcs4--0\farcs6).  Observations yield a
limiting magnitude of 22.7 ($J$) and 22.6 ($H$; 3$\sigma$ confidence
level). {\tt SOFI} $Ks$ images were obtained on 2002 February 19
(seeing 0\farcs8). The exposure time was of 3000\,s, and a limiting
magnitude of $Ks$$\sim$21.4 was reached. In all these cases, single
5\,s-long exposure images were taken for each filter with offsets of
40\arcsec\ in order to sample and subtract the variable IR background.
Finally, $H$ and $K'$ images were obtained with {\tt AOB} on 2002
August 17 with effective exposure time of 2700\,s (for each filter;
seeing of 0\farcs4--0\farcs5). Thanks to the adaptive optics we
obtained a source PSF of $\sim$0\farcs12 yielding a limiting magnitude
of 23.1 ($H$) and 21.8 ($K'$). During the latter run single 60\,s-long
exposure images were taken with random offsets in the
4\arcsec--10\arcsec\ range.  Standard IR software packages were used
for sky frame subtraction and image coaddition ({\tt Eclipse} and {\tt
IRDR}; Devillard 1997 and Sabbey et al. 2001).

To register the \AXAF\ coordinates of \src\ on our optical/IR images,
we computed the image astrometry using, as a reference, the positions
of stars
\begin{inlinefigure}
\psfig{figure=v_vFvnewe.ps,height=8.9cm} 
\caption{Broad band energy spectrum of \src.  X--ray raw data are
taken from the medium and low energy instruments on board \BSAX\
(filled squares) while the solid upper curves are the unabsorbed
fluxes for the black body (BB), the power law (PL), and the sum of the
two components. On the lower left corner of the plot are the
optical/IR fluxes: plus signs are absorbed values, while filled
triangles and circles are unabsorbed ones for $A_{\rm V}$=7.8 and
11.2, respectively. The thick dotted, dash--dotted and solid lines are
the fossil disk models discussed in the text.}
\end{inlinefigure}

\noindent (about 20 for ESO instruments) selected from the GSC2.2
catalog which has an intrinsic absolute accuracy of about
0\farcs2--0\farcs4 (depending on magnitude and sky position of
stars).
After taking into account the uncertainties of the source X--ray
coordinates (0\farcs7), the rms error of our astrometry (0\farcs12)
and the propagation of the intrinsic absolute uncertainties on the
GSC2.2 coordinates (we assumed a value of 0\farcs3), we estimated the
final accuracy to be attached to the \src\ position of about
0\farcs8. Figure\,1 shows a region of 28\arcsec$\times$20\arcsec\
around the \src\ position in the $K'$ band CFHT image (90\% confidence
level \R\ and \AXAF\ uncertainty circles superimposed). In the inset
we show the close--up of the region around the \AXAF\ position with
the counterpart candidates marked.

The 3.6-m ESO optical $R$ band data show that no object is consistent
with the \AXAF\ position of \src\ down to a limiting magnitude of
about 26.5.  On the other hand, two relatively faint objects are
detected in the IR images carried out at the CFHT (marked with A and B
in Figure\,1). The latter two sources have the following IR magnitudes
as inferred from CFHT (NTT) data: $K'$=17.53$\pm$0.02,
$H$=18.85$\pm$0.05 ($Ks$=18.3$\pm$0.1, $H$=18.6$\pm$0.1 and
$J$=20.9$\pm$0.1) for object A, $K'$=20.00$\pm$0.08 and
$H$=20.43$\pm$0.07 for B. Unfortunately, no $J$ measurement is
available for object B due to its faintness and likely confusion with
object A in SOFI images.

Figure\,2 shows the color--color diagram of the objects in the field
with star A marked.  Note that the hydrogen column density to \src,
inferred from the X--ray observations is $N_{\rm
H}$=1.40(6)$\times$10$^{22}$ cm$^{-2}$ (Israel et al. 2001, Rea et
al. 2003), converts into a visual extinction of $A_V\simeq 7.8$ for a
typical dust--to--gas ratio (Predehl \& Schmitt 1995, see thick solid
line in Figure\,2). In the $J$--$K'$ versus $J$--$H$ plane the
reddening vector lies roughly parallel to the locus of points
belonging to star theoretical isochrones at solar chemical composition
and stellar ages ranging from 10\,Myr to 8\,Gyr (Bono et al. 1997,
2000; a value of 3.1 has been assumed for the parameter
R=$A_V$/$E(B$-$V)$; Cardelli et al. 1989, Fitzpatrick 1999).  We note
that the synthetic sequences account for nearly all the objects but
few red stars ($J$--$K'$ colors larger than 3) that are uniformly
distributed over the FOV. This is likely due to the parameter R, the
value of which is likely larger than that assumed as the ``nominal''
one (in particular when looking at several spiral arms through the
Galactic plane, as in the case of \src).  Finally, we used the
information on the colors of the known IR counterparts to AXPs
($H$--$K'$=1.1 and 1.4 for \uu\ and \car, respectively; Israel et
al. 2003; Wang \& Chakrabarty 2002), to identify the most probable
region of the diagram where the IR counterpart to \src\ would lie, if
similar colors apply and assuming $A_V$$\geq$6.5. Object A is clearly
included in the above region and it is also the only object showing
unusual colors with respect to both the theoretical star sequences and
all other detected objects. All the above results make object A a
robust candidate, and in the following we will refer to it as the
likely counterpart to \src. Based again on the color, we regard
candidate B as a far less likely candidate ($H$--$K'$=0.4).

\section{DISCUSSION}

The proposed IR counterpart to \src\ is by far the brightest one among
those of AXPs: $K'$=21.7, $K'$=20.0, $Ks$=19.4 for \ee, \uu, and \car,
respectively (Hulleman et al. 2001 and 2002, Israel et al. 2003, Wang
\& Chakrabarty 2002). We note that the relatively large X--ray-to-IR
unabsorbed flux ratio of about 500 is lower than that obtained for
other AXPs. However, \src\ is also the most luminous AXP in the
0.5--10\,keV band with an unabsorbed luminosity of $\geq$
4$\times$10$^{35}$ erg\,s$^{-1}$ (assuming a distance $\geq$5\,kpc).
In order to characterise the broadband energy distribution of the
source, we plotted the IR through X--ray data of \src\ in
Figure\,3. X--ray fluxes have been inferred by using the
phase--averaged spectral parameters obtained with \BSAX\ data (Israel
et al.  2001, Rea et al. 2003). Two different values of the extinction
have been used for model fitting purposes: $A_V$=7.8 and 11.2
(triangles and circles, respectively), corresponding to the N$_H$
inferred from the X--ray spectra and the Galactic absorption in the
direction of the source, respectively.

The blackbody component detected in the X--rays ($kT$$\sim$0.45 keV;
Israel et al. 2001, Rea et al. 2003) cannot account for the
(relatively high) IR flux.  Similarly, the power--law component, if
extrapolated to IR wavelengths, would imply a much higher IR flux
level which is simply not observed; a cut--off must therefore occur
somewhere in the UV/optical bands.  Regardless of the exact position
of the cut--off and its origin, we note that the X--ray-to-IR emission
of \src\ cannot be fit with a simple spectral component (similar to
the case of \uu\ and \ee; Hulleman et al. 2000, 2001). 

{\rc The magnetar model does not account for the observed IR emission
or IR variability of AXPs and no predictions can be verified.
Therefore, we discuss the above observational findings in the context
of models based on fossil disks (Chatterjee et al. 2000; Alpar 2001)
since they make clear predictions for the IR emission. However, up to
now the latter models are unable to account for the bursts seen in
AXPs and SGRs.}. In the case of a fossil disk the IR emission arises
mainly from two components: viscous dissipation and reprocessing of
X--ray flux impinging on the disk from the pulsar (Perna et al. 2000,
Perna \& Hernquist 2000, Alpar 2001).  Specifically, at IR wavelengths
and based on the X--ray luminosity of \src, the dominant emission
component is expected to be X--ray irradiation. The models we
considered have been computed for a distance of 5\,kpc {\rc and the
observed X--ray luminosity as an input parameter}, but since the disk
luminosity scales almost linearly with the X--ray luminosity, the
result is nearly independent of the distance. {\rc As far as
the X--ray spectrum, in the accretion model (with magnetic fields of
$\sim$10$^{12}$ Gauss) one would expect the emission to be roughly a
blackbody produced in a region (polar cap) much smaller than the area
of the star.  Current X--ray data are consistent with this, but not
good enough to allow a firm discrimination with respect to the X--ray
spectrum of a magnetar, in which case one would expect the area of the
emitting region to be consistent with the whole surface of the star
(Perna et al. 2001).}

In Figure\,3 we show disk models obtained with the inner and outer
radius of the disk, $r_{in}$ and $r_{out}$, and the inclination left
free to vary (however, note that the inclination only scales the
fluxes, while $r_{in}$ and $r_{out}$ modify the shape of the model and
the flux peak position; see Perna et al. 2000 for details). The best
fit was obtained for a small truncated disk, with $r_{in}$ and
$r_{out}$ equal to 10$^{11}$ and 5$\times$10$^{11}$\,cm, respectively
(dotted line in Figure\,3). However, this solution would not be able
to explain the X--ray emission from AXPs, being the value of $r_{in}$
larger than the corotation radius ($\sim$8$\times$10$^9$\,cm), the
characteristic value of $r_{in}$ in order to have accretion on the
neutron star surface, and at the same time, spin--down.
Therefore, we fixed $r_{in}$ to the corotation radius for two
different values of extinction. The results of the fit are shown in
Figure\,3 (dashed--dotted and solid lines, respectively). Values of
$r_{out}$=10$^{12}$ and 8$\times$10$^{11}$\,cm were obtained for
$A_V$=7.8 and 11.2, respectively, (fallback compact disks; Menou et
al. 2001). In both cases the $J$ measurement is lower than the model
expectations (2$\sigma$).  The above solution might be applied also to
the case of an accretion disk formed by mass transfer from a (light)
companion, although in this case an (unknown) amount of IR flux should
be originated also from the star surface. We note that IR flattening
has been recently discovered also for \uu, and likely during an X--ray
bursting activity phase of \car\ (Wang \& Chakrabarty 2002, Israel et
al. 2003). The IR flattening or excess might represent a new important
property shared by AXPs.  

The IR emission of AXPs in the magnetar scenario, remains to be
addressed through detailed modeling.  However, we note that the
presence of a disk would not be necessarily in contrast with the
magnetar scenario.  Three AXPs are associated with supernova remnants
and likely embedded in dense regions.  Therefore a ``hybrid'' scenario
in which a magnetar irradiates a fossil disk (the matter of which is
not accreting in this case) might not be unreasonable, though it
should be confirmed through additional future observations and its
consistency checked through theoretical studies.  

{\rc In conclusion, even though we propose a possible solution for the
IR excess observed in AXPs, we note that none of the proposed
theoretical models (at least in their present form) seem to be able to
easily account simultaneously for the IR, optical and X--ray emission
of AXPs. }

 
\acknowledgments
This work is supported through CNAA, ASI, CNR and
Ministero dell'Universit\`a e Ricerca Scientifica e Tecnologica
(MURST--COFIN) grants. The authors thank Olivier Hainaut, Leonardo
Vanzi and the NTT Team for their kind help during ESO observations. We
also thank Olivier Lai and the CFHT team for their support. We finally
thanks Hank Donnelly of the \AXAF\ Team.

\end{document}